\documentclass[aps,prl,twocolumn,10pt,showpacs,superscriptaddress,amsmath,amssymb,nobibnotes]{revtex4-1}
\usepackage{graphicx}
\usepackage{amsfonts}    
\usepackage{graphicx}   
\usepackage{verbatim}   
\usepackage{color}      
\usepackage{subfigure}  
\usepackage{hyperref}   
\usepackage{natbib}
\usepackage{booktabs}
\usepackage{threeparttable}
\usepackage{dcolumn}
\usepackage{multirow}
\usepackage{float}
\usepackage{ulem}
\usepackage{bm}
\usepackage{algorithm}
\usepackage{algorithmicx}
\usepackage[noend]{algpseudocode}
\usepackage{CJK}

\makeatletter
\def\BState{\State\hskip-\ALG@thistlm}
\makeatother
\begin{document}
\begin{CJK*}{UTF8}{gbsn}


\title{Proposal for Sequential Stern-Gerlach Experiment with Programmable Quantum Processors}

\author{Meng-Jun Hu (\CJKfamily{gbsn}胡孟军)}
\email{humj@baqis.ac.cn}
\affiliation{Beijing Academy of Quantum Information Sciences, Beijing 100193, China}

\author{Haixing Miao (\CJKfamily{gbsn}缪海兴)} 
\affiliation{State Key Laboratory of Low Dimensional Quantum Physics, Department of Physics, Tsinghua University, Beijing, China}

\author{Yong-Sheng Zhang (\CJKfamily{gbsn}张永生)}
\affiliation{Laboratory of Quantum Information, University of Science and Technology of China, Hefei 230026, China }%
\affiliation{Synergetic Innovation Center of Quantum Information and Quantum Physics,
University of Science and Technology of China, Hefei 230026, China}
\affiliation{Hefei National Laboratory, University of Science and Technology of China, Hefei, 230088, China}

\date{\today}

\begin{abstract}
The historical significance of the Stern-Gerlach experiment lies in its provision of the initial evidence for space quantization. Over time, its sequential form has evolved into an elegant paradigm that effectively illustrates the fundamental principles of quantum theory. To date, the practical implementation of the sequential Stern-Gerlach experiment has not been fully achieved. In this study, we demonstrate the capability of programmable quantum processors to simulate the sequential Stern-Gerlach experiment. The specific parametric shallow quantum circuits, which are suitable for the limitations of current noisy quantum hardware, are given to replicate the functionality of Stern-Gerlach devices with the ability to perform measurements in different directions. Surprisingly, it has been demonstrated that Wigner's Stern-Gerlach interferometer can be readily implemented in our sequential quantum circuit. With the utilization of the identical circuits, it is also feasible to implement Wheeler's delayed-choice experiment.
We propose the utilization of cross-shaped programmable quantum processors to showcase sequential experiments, and the simulation results demonstrate a strong alignment with theoretical predictions. With the rapid advancement of cloud-based quantum computing, such as BAQIS Quafu, it is our belief that the proposed solution is well-suited for deployment on the cloud, allowing for public accessibility.
Our findings not only expand the potential applications of quantum computers, but also contribute to a deeper comprehension of the fundamental principles underlying quantum theory.
\\


\end{abstract}

\maketitle
\end{CJK*}


\section{1. Introduction}
\label{sec:intro}

The Stern-Gerlach (S-G) experiment, which was conducted in 1922, served as the initial demonstration of the quantization of the magnetic moment \cite{SG1, SG2, SG3}. Furthermore, this experiment played a pivotal role in paving the way for modern atomic physics \cite{SG4}. Interestingly, it was only after the discovery of spin \cite{spin0} that it became apparent the S-G setup was actually measuring the electron spin \cite{spin1}. Since spin represents the most basic form of a two-level quantum system, the S-G experiment and its sequential variant have been extensively employed to elucidate the fundamental principles of quantum mechanics, e.g., statevector, quantum superposition, and quantum entanglement \cite{book1, book2, book3}. A spin that traverses two S-G devices consecutively, each with different measurement directions as depicted in Fig. 1a, would yield distinct outcomes depending on the alteration of the measurement order. This significant deviation from classical theory serves as a clear demonstration of the fundamental feature of quantum theory, namely, the non-commutation of certain observables \cite{Dirac}. The well-known Heisenberg uncertainty relation arises from the non-commutativity of particle position and momentum \cite{Heisenberg}. In addition to the conventional S-G setup, Wigner has proposed an alternative configuration in which two spin paths are combined to restore coherence, as illustrated in Fig. 1b. This arrangement aims to elucidate the difference of entanglement state and mixed state in the context of measurement \cite{Wigner}. The variant configuration referred to as the S-G interferometer \cite{Schwinger1, Schwinger2, Schwinger3} has garnered significant interest in recent times due to its potential in quantum gravity testing \cite{Gravity}. Therefore, the implementation of sequential S-G experiments holds immense importance in enhancing our comprehension of quantum theory.

\begin{figure}[tp]
\includegraphics[scale=0.3]{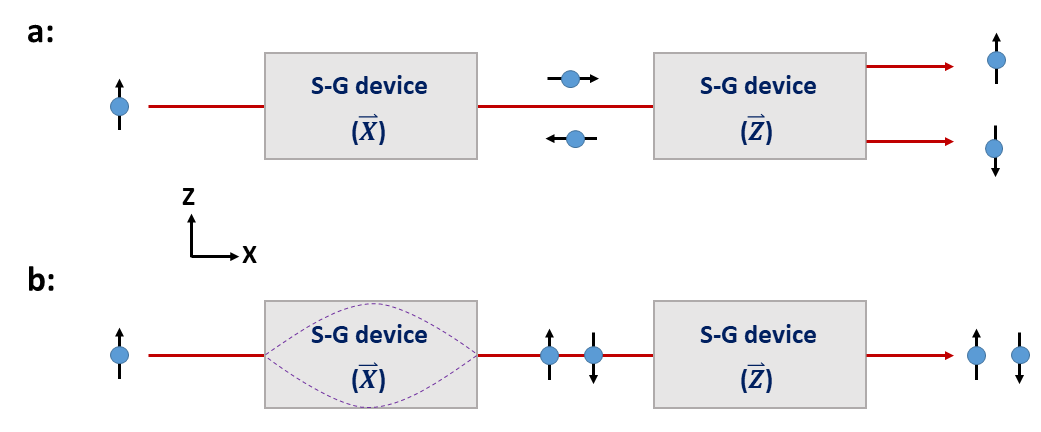}
\caption{ A schematic diagram illustrating sequential Stern-Gerlach experiments. {\bf a:} Sequential spin measurements are conducted using two S-G setups, each with different measurement directions;  {\bf d:} Wigner's S-G interferometer that recombine two spin paths after the initial S-G device. }
\label{fig1l}
\end{figure} 

\begin{figure*}[t]
\includegraphics[scale=0.45]{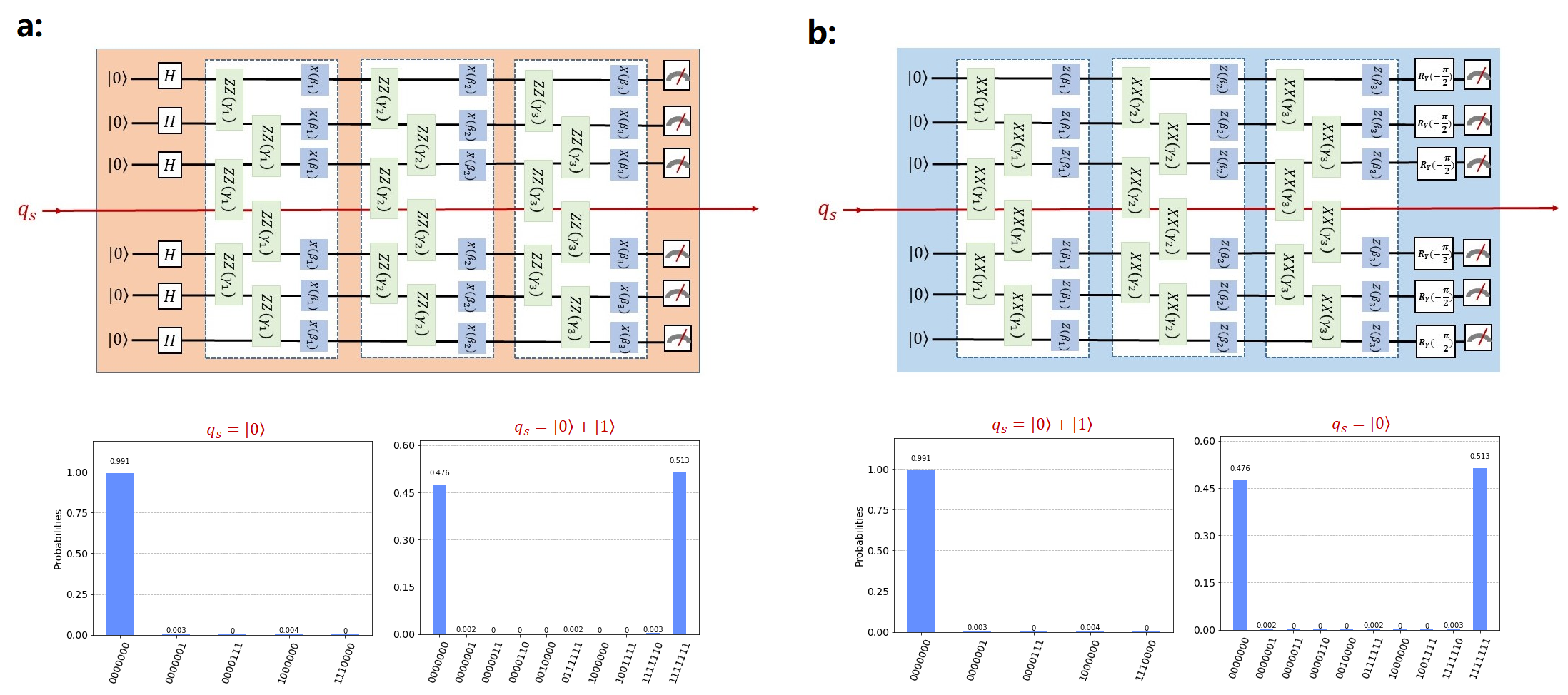}
\caption{Quantum circuit realization of Stern-Gerlach devices. {\bf a:} The S-G quantum circuit in the $\hat{\sigma}_{Z}$ basis; {\bf b:} The S-G quantum circuit in the $\hat{\sigma}_{X}$ basis. The optimal parameters $(\vec{\gamma}^{*}, \vec{\beta}^{*})$ in quantum circuits can be obtained by variational optimization with $q_{s}$ being the definite state, e.g., $|0\rangle$. The second row dosplays the simulation results of two S-G quantum circuits with the same optimal parameters. Note that the two quantum circuits are measured in different basis, and the result $\lbrace 0, 1\rbrace$ on each qubit corresponds to eigenstates in the measurement basis. }
\label{fig:circuit}
\end{figure*}

Although the S-G setup has already been extensively utilized in previous studies \cite{use1, use2, use3}, the implementation of the sequential S-G setup is not as straightforward as it may appear. Extremely precise calibration and control are necessary, posing a challenge to achieve even in contemporary times. Fortunately, the recent advancements in programmable quantum processors \cite{QPU1, QPU2, QPU3, QPU4, QPU5, QPU6, QPU7, QPU8, QPU9, QPU10} have facilitated the possibility of employing an alternative approach to showcase sequential S-G experiments. While currently only noisy intermediate-scale quantum (NISQ) processors \cite{NISQ} are accessible, they are deemed adequate for our intended objectives. In this study, we demonstrate the implementation of S-G devices with varying measurement orientations through the utilization of parametric quantum circuits. Similar to the existing variational quantum algorithms \cite{vqe1, vqe2, vqe3}, the classical optimizer is employed to determine the optimal parameters of quantum circuits \cite{Qcover}. The quantum circuits, when configured with optimal parameters, will function as S-G devices for the purposes of spin measurement, specifically in the context of qubits. The variational shallow quantum circuit consists solely of nearest-neighbor two-qubit gates, which makes our S-G quantum circuits renders them highly compatible with NISQ hardware implementation. In order to enhance the performance of sequential S-G experiment, we suggest utilizing cross-shaped quantum processors for executing S-G quantum circuits. 

\label{sec:model}
\begin{figure*}[t]
\includegraphics[scale=0.44]{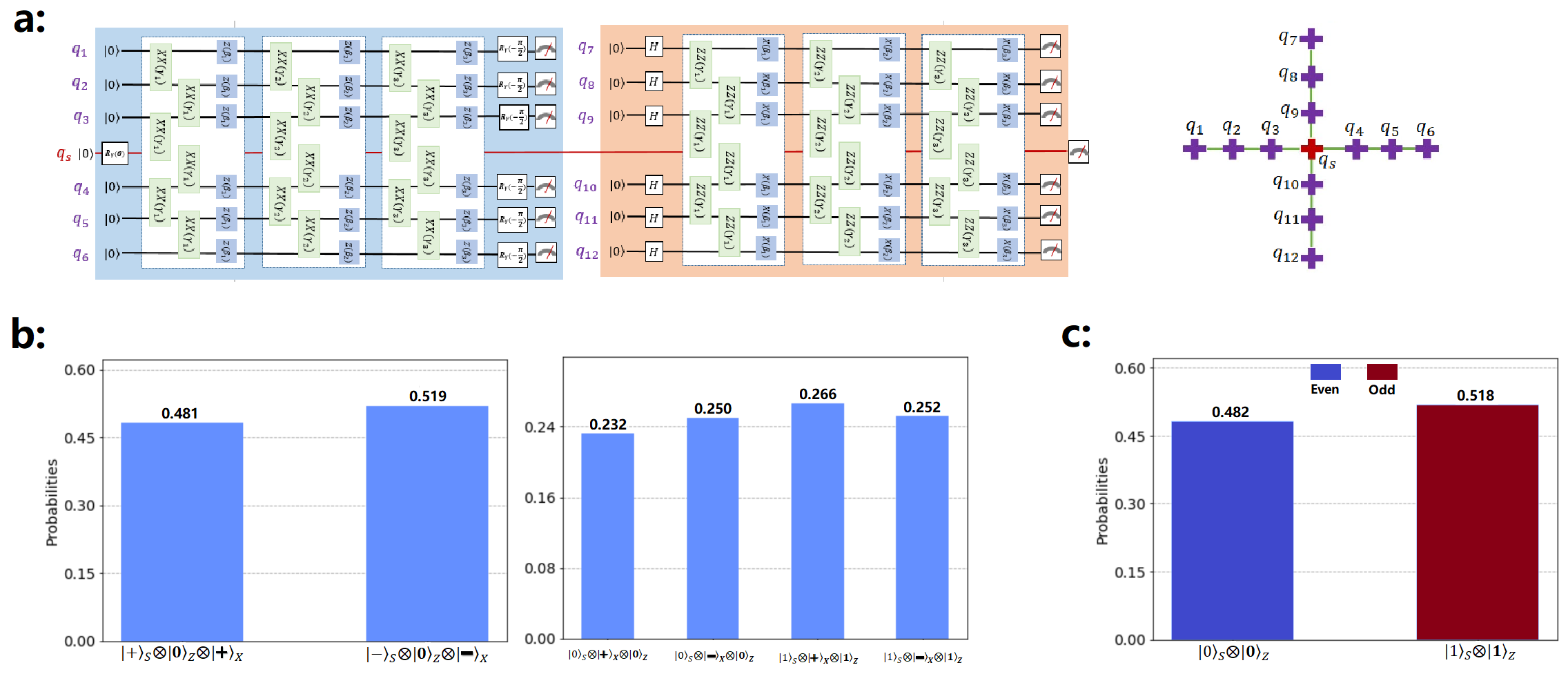}
\caption{{\bf a:} Quantum circuit realization of sequential Stern-Gerlach experiment based on cross-shaped quantum processors. The Wigner's S-G interferometer can be demonstrated by just removing all the rotation gates $R_{y}(-\frac{\pi}{2})$. {\bf b:} Simulation results with the qubit $q_{S}=|0\rangle_{S}$ in sequential S-G quantum circuit. The first is the measurement result of $\hat{\sigma}_{Z}$ first and then $\hat{\sigma}_{X}$, while the second is the measurement result of $\hat{\sigma}_{X}$ first and then $\hat{\sigma}_{Z}$. {\bf c:} Result of the Wigner's S-G interferometer with $q_{S}=|0\rangle_{S}$. The blue one implies that the resulting states of probe qubits in the first S-G device of $\hat{\sigma}_{X}$ measurement are even states, while the red one implies that the resulting states are odd states. Even/Odd state refers to there are even/odd probe qubits in the state $|1\rangle$. The collective states, e.g., $|\bm 0\rangle_{Z}$ means that majority of probe qubits of $\hat{\sigma}_{Z}$ measurement are in the state $|0\rangle$. }
\label{fig:result}
\end{figure*}

\section{2. Quantum Circuit Realization of Stern-Gerlach Device}

According to the quantum measurement theory, the spin of a particle is entangled with its path in the S-G device before ultimately collapsing into one of its eigenstates upon interaction with the screen \cite{theory}. Since the spin state is destroyed upon measurement, it becomes unfeasible to conduct subsequent measurements. Therefore, it is preferable to employ non-destructive measurement techniques, wherein the state of spin can be inferred from the measurement of probe without causing any damage to the system. In the context of a quantum circuit analogy, the concept of spin is substituted with a qubit, while additional qubits within the circuit serve as probes to deduce {\it which-way} information pertaining to the spin. The gate operations within the quantum circuit establish entanglement between the qubit and the probe. At the end of the quantum circuit, measurements are conducted on the probe in order to deduce the state of the qubit, thereby achieving a non-destructive measurement. The crucial aspect is to devise a quantum circuit that effectively entangles the qubit and the probe, resulting in the creation of a cat-like state. For instance, when the measurement is conducted on the $\hat{\sigma}_{Z}$ basis, we expected that the final state of the circuit, for any arbitrary qubit state expressed in the $\hat{\sigma}_{Z}$ basis $a|0\rangle_{S}+b|1\rangle_{S}$ with $|a|^{2}+|b|^{2}=1$, be 
\begin{equation}
|\Psi\rangle_{f}= a|0\rangle_{S}|\bm 0\rangle_{P}+b|1\rangle_{S}|\bm 1\rangle_{P}
\end{equation}
with collective states $|\bm 0\rangle_{P}$ and $|\bm 1\rangle_{P}$ represent that the majority of probe qubits are in states $|0\rangle$ and $|1\rangle$, respectively. Similarly, the measurement is conducted on the $\hat{\sigma}_{X}$ basis, the resultant final state of the measurement circuit should be represented as $|\Psi\rangle_{f}=a|+\rangle_{S}|\bm +\rangle_{P}+b|-\rangle_{S}|\bm -\rangle_{P}$, where $|\pm\rangle=(|0\rangle\pm |1\rangle)/\sqrt{2}$ denotes the eigenstates of $\hat{\sigma}_{X}$.

The aforementioned S-G quantum circuits can be implemented using parametric quantum circuits, as depicted in Fig. 2. In our previous study \cite{Hu}, the quantum circuit of Fig. 2a has been constructed to generate a cat-like state in the $\hat{\sigma}_{Z}$ basis, as described by Eq. (1). Similar to the Stern-Gerlach setup that is utilized for spin measurement, our parametric quantum circuits, with their specially designed circuit structure, serve as a means of qubit measurement. The process of optimizing circuit parameters can be likened to the calibration process of the Stern-Gerlach setup prior to conducting  official measurements. Once the optimal parameters has been established in the circuit, any qubit in the state $a|0\rangle+b|1\rangle$ that is input into the circuit will produce the cat-like state, as exemplified by Eq. (1). This cat-like state allows for the extraction of qubit information through the measurement of probe qubits. This process bears resemblance to the Stern-Gerlach setup, which involves the entanglement of spin with its corresponding which-way path, described as $a|\uparrow\rangle|up\rangle+b|\downarrow\rangle|down\rangle$.
The circuit is initially prepared in the state $|\psi_{0}\rangle=(a|0\rangle_{S}+b|1\rangle_{S})\otimes |+\rangle^{\otimes 2N}$ by applying Hadamard gates on probe qubits, where $N=3$ in this particular case. The selection of $2N$ probe qubits in this study is primarily motivated by two key factors. The first issue to consider is the presence of measurement errors, which can be mitigated by employing multiple probe qubits to minimize the occurrence of misreading to a negligible level. The second issue pertains to the current NISQ devices, which are plagued by noisy operations. Consequently, it is imperative to limit the number of qubits in order to ensure accurate and effective calculations.
The evolutionary component of the circuit comprises $m$ layers of similar gate operations including a total of $2m$ parameters. Each layer is composed of parametric two-qubit gates $ZZ(\gamma)=e^{i\gamma\hat{\sigma}_{Z}\hat{\sigma}_{Z}}$ applied to the nearest qubits, as well as single-qubit rotation gate $X(\beta)=e^{i\beta\hat{\sigma}_{X}}$ applied to the probe qubits. The ultimate state of the parametric circuit will be
\begin{equation}
|\vec{\gamma}, \vec{\beta}\rangle=\prod X(\beta_{m})\prod ZZ(\gamma_{m})\cdots\prod X(\beta_{1})\prod ZZ(\gamma_{1})|\psi_{0}\rangle
\end{equation}
with $\vec{\gamma}=\lbrace\gamma_{1}, \cdots, \gamma_{m}\rbrace$ and $\vec{\beta}=\lbrace\beta_{1}, \cdots, \beta_{m}\rbrace$. In principle, increasing the number of $m$ leads to improved output results of the optimized circuit. However, current NISQ processors only support low depth circuit for effective calculations. 
We observe that the value of $m=3$ is sufficient in the case where $N=3$, as demonstrated in Fig. 2 through simulation results.
The circuit depth is limited to the application of only nearest-neighbor two-qubit gates, which enhances the feasibility of implementing our circuit on existing NISQ hardware.
To implement the S-G quantum circuit in the $\hat{\sigma}_{Z}$ basis, it is necessary to determine the optimal parameters $(\vec{\gamma}^{*}, \vec{\beta}^{*})$ such that $|\vec{\gamma}^{*}, \vec{\beta}^{*}\rangle=|\Psi\rangle_{f}$. The determination of the optimal parameters $(\vec{\gamma}^{*}, \vec{\beta}^{*})$ can be achieved through the application of the variational method, utilizing the cost function $\langle\vec{\gamma}, \vec{\beta}|\hat{H}|\vec{\gamma}, \vec{\beta}\rangle$, in which $\hat{H}=-\sum_{k=1}^{2N}\hat{\sigma}_{Z}^{k}\hat{\sigma}_{Z}^{k+1}$ is the Ising Hamiltonian with nearest-neighbor interactions. The minimization of cost function output the desired optimal parameters $(\vec{\gamma}^{*}, \vec{\beta}^{*})$. Since the lowest energy value of $\hat{H}$ is known, the closeness of optimized cost function to this value reflects the quality of cat-like state preparation.
This procedure bears resemblance to the calibration of the device before official measurements.
Specifically, the initial state of the measurement qubit $q_{s}$ is set to $|0\rangle$, followed by the execution of the circuit to calculate the $\langle\vec{\gamma}, \vec{\beta}|\hat{H}|\vec{\gamma}, \vec{\beta}\rangle$ using randomly assigned initial parameters. The classical optimizer, such as COBYLA, is utilized to minimize the cost function, and the optimal parameters is obtained once the convergence condition has been met. The optimal parameters $(\vec{\gamma}^{*}, \vec{\beta}^{*})$ are adjusted in the circuit, thereby completing the S-G quantum circuit.

The S-G quantum circuit of the $\hat{\sigma}_{X}$ basis, as depicted in Fig. 2b, is formed by substituting $ZZ(\gamma)$ with $XX(\gamma)=e^{i\gamma\hat{\sigma}_{X}\hat{\sigma}_{X}}$, and $X(\beta)$ with $Z(\beta)=e^{i\beta\hat{\sigma}_{Z}}$. We have excluded the Hadamard gates from the circuit as $|0\rangle$ is already in a superposition state on the $\hat{\sigma}_{X}$ basis. Since the measurement of qubits in a quantum circuit only yields results in the form of $\lbrace|0\rangle, |1\rangle\rbrace$, rotation gates $R_{y}(-\frac{\pi}{2})$ are applied to the probe qubits prior to measurement in order to convert the state $|+\rangle$ to $|0\rangle$ and the state $|-\rangle$ to $|1\rangle$, respectively. We aim to simulate the two S-G quantum circuits with different basis using the same optimal parameters $(\vec{\gamma}^{*}, \vec{\beta}^{*})$. The results depicted in Fig. 2 align closely with the theoretical predictions. Interestingly, the resulting states of the two circuits have the same probability distribution on their own basis. This assertion appears to be reasonable due to the S-G setup along the $X$ direction can be achieved by simply rotating the identical setup along the $Z$ direction.
In practical application, however, the optimization is contingent upon the performance of the NISQ hardware and the characteristic of the specific quantum circuit. 

In summary, we have put forth a novel variational quantum circuit specifically designed for the current NISQ processors to replicate the functionality of S-G setup. In the S-G setup, the spin that to be measured becomes entangled with its which-way path described as $a|\uparrow\rangle|up\rangle+b|\downarrow\rangle|down\rangle$, where the extraction of which-way information provides us the information of spin state. In the context of the quantum circuit analog, the qubit is utilized to simulate the spin, while the role of the which-way path is replaced by probe qubits that are entangled with the qubit to be measured. The final state of the variational quantum circuit with optimized parameters exhibit a cat-like state represented as $a|0\rangle|\bm 0\rangle+b|1\rangle|\bm 1\rangle$, as described in Eq. (1), in order to successfully simulate the desired outcome of the measurement process in the Stern-Gerlach setup. In order to ensure the desired cat-like state in the final circuit, We select the cost Hamiltonian for the variational quantum circuit as the Ising Hamiltonian with nearest-neighbor interactions. This is due to the fact that one of ground states of the cost Hamiltonian is the GHZ state, represented as $a|0\cdots0\rangle+b|1\cdots1\rangle$. The circuit structure depicted in Fig. 2 ensures the optimal arrangement of layer, thereby minimizing the cost Hamiltonian. The optimized circuit would result in a cat-like state at the end. The number of parameters in the circuit is determined by how close we want the value of cost function $\langle\vec{\gamma}, \vec{\beta}|\hat{H}|\vec{\gamma}, \vec{\beta}\rangle$ approach the ground state eigenvalue of $\hat{H}$. 
 
\section{3. Sequential Stern-Gerlach Experiment Realization }
With the availability of S-G quantum circuits employing different measurement basis, we are now prepared to delve into the implementation of sequential S-G experiments.  The quantum circuit depicted in Fig. 3a illustrates the implementation of the sequential S-G experiment. In this experiment, a qubit $q_{S}$ initially traverses the S-G quantum circuit in the $\hat{\sigma}_{X}$ basis, and subsequently enters the S-G quantum circuit in the $\hat{\sigma}_{Z}$ basis. This sequential S-G quantum circuit, which performs a $\hat{\sigma}_{X}$ measurement followed by a $\hat{\sigma}_{Z}$ measurement on the qubit $q_{S}$, can be considered as a quantum analog of the sequential spin measurement depicted in Fig. 1a. The sequential measurement of the first $\hat{\sigma}_{Z}$ measurement and then the $\hat{\sigma}_{X}$ measurement of $q_{S}$ can be easily achieved by modifying the S-G quantum circuits component.  
The sequential S-G quantum circuit is best executed on a cross-shaped quantum processor, as depicted in the right portion of Fig. 3a, with the qubit $q_{S}$ positioned at the center. The qubits located in the horizontal and vertical lines are utilized for the implementation of S-G quantum circuits in the $\hat{\sigma}_{X}$ basis and the $\hat{\sigma}_{Z}$ basis, respectively. In practical applications, it is necessary to conduct on-chip optimization in order to determine the optimal parameters for two S-G quantum circuits prior to conducting the experiment. With the utilization of cross-shaped processors, the alteration in the measurement sequence of non-commuting observables of $q_{S}$ can be achieved by simply modifying the execution order of the horizontal and vertical qubits. 

We are now able to engage in a comprehensive discussion on the implementation of Wigner's S-G interferometer, which effectively restores spin coherence subsequent to the initial S-G device. It is, nonetheless, not immediately apparent upon initial observation. We will demonstrate that this can be accomplished effortlessly by simply altering the measurement basis of the initial S-G quantum circuit. To be concrete, we consider the qubit $q_{S}$ initially prepared in state $|0\rangle_{S}$ passing through a sequential S-G quantum circuit, as depicted in Fig. 3. Denoting the initial states of the two S-G quantum setup as $|\bm 0\rangle_X$ and $|\bm 0\rangle_Z$ respectively, we obtain the final state of the whole system at the end of the circuit as
\begin{equation}
|\Phi\rangle_{F}=\frac{1}{2}[|0\rangle_{S}(|\bm +\rangle_{X}+|\bm -\rangle_{X})|\bm 0\rangle_{Z}+|1\rangle_{S}(|\bm +\rangle_{X}-|\bm -\rangle_{X})|\bm 1\rangle_{Z}].
\end{equation}
If the probe qubits($q_{1}-q_{6}$) in the initial S-G quantum circuit can be projected into the states $|\bm +\rangle_{X}\pm|\bm -\rangle_{X}$, then the qubit $q_{S}$ will undergo a collapse and be in either the state $|0\rangle_{S}$ or $|1\rangle_{S}$. In this particular instance, the Wigner's interferometer analogy has been implemented.
The measurement of the second S-G quantum device in the $\hat{\sigma}_{Z}$ basis will not cause any changes to the state of $q_{S}$. It is, however, challenging to accurately project the $|\bm +\rangle_{X}\pm|\bm -\rangle_{X}$ states onto the probe qubits. To address this challenge, it is important to acknowledge that the state$|\bm +\rangle_{X}+|\bm -\rangle_{X}$ exclusively consists of even states $\lbrace|0\rangle, |2\rangle, |4\rangle, |6\rangle\rbrace$, while the state $|\bm +\rangle_{X}-|\bm -\rangle_{X}$ comprises odd states $\lbrace|1\rangle, |3\rangle, |5\rangle\rbrace$. Here the notation $|n\rangle$ represents the number of probe qubits in the state $|1\rangle_{X}$. This state is referred to as the even/odd state, depending on whether $n$ is even/odd. It is now evident that the measurement of probe qubits on the $\hat{\sigma}_{Z}$ basis is sufficient, which involves the removal of all rotation gates $R_{y}(-\frac{\pi}{2})$. If the results is either even or odd state, then the qubit $q_{s}$ will collapsed into the $|0\rangle_{S}$ or $|1\rangle_{S}$ state. The realization of of Wigner's S-G interferometer using the original S-G device is surprisingly achievable with the quantum circuit version.

The sequential S-G quantum circuit, consisting of a total of $13$ qubits, as depicted in Fig. 3a, is simulated in this study by utilizing the IBM {\it Qiskit} package \cite{qiskit}. The optimal parameters $(\vec{\gamma}^{*}, \vec{\beta}^{*})$ are predetermined in the circuit are obtained in advance using the BAQIS {\it Qcover} package \cite{Baqis}. Our simulations are based on the assumption that all qubits and gate operations are ideal, meaning that no circuit noise models are incorporated. 
The qubit $q_{S}$, which emulates the spin to be measured as depicted in Fig. 1, is initially prepared in the state $|0\rangle_{S}$. Fig. 2b displays the simulation results of sequentially measuring the qubit observables $\hat{\sigma}_{Z}$ and $\hat{\sigma}_{X}$. The left panel of Fig. 3b displays the outcome of measuring $\hat{\sigma}_{Z}$ prior to $\hat{\sigma}_{X}$, whereas the right panel shows the outcome of measuring $\hat{\sigma}_{X}$ prior to $\hat{\sigma}_{Z}$. The significant disparity observed when altering the sequence of measurement serves as a clear demonstration of the non-commutative nature of the observables $\hat{\sigma}_{Z}$ and $\hat{\sigma}_{X}$. In the outcome presented in the left panel, it is evident that there exists a possibility for the state of $q_{S}$ to be completely distinct from its original state after the measurement process concludes. The explanation of this phenomenon does not conform to classical principles, but rather arises as a natural outcome of the superposition of quantum states. Fig. 3c illustrates the simulation outcome of Wigner's S-G interferometer, achieved by eliminating all rotation gates $R_{y}(-\frac{\pi}{2})$ in the sequential quantum circuit. The sampling results of the circuit can be categorized into two groups. The blue category indicates that the resulting states of probe qubits are even states, while the red category indicates that the resulting states are odd states. It is evidence from the results that $q_{S}$ is restored to the state $|0\rangle_{S}$, which represents a coherent state in the basis of $\hat{\sigma}_{X}$ for even states. Conversely, for odd state, $q_{S}$ is restored to the state $|1\rangle_{S}$.

\section{4. Discussion and Conclusion}
\label{sec:conclusion}
The previous proposed sequential S-G quantum circuit for Wigner's S-G interferometer can also serve as a demonstration of wave-particle duality. This concept is best exemplified by Wheeler's delayed-choice gedanken experiment \cite{Wheeler}.  In the delayed-choice experiment, the option to open or close the second beam splitter $\mathrm{BS_{out}}$ within the Mach-Zehnder interferometer is available after a single photon has traversed the first beamsplitter. When  $\mathrm{BS_{out}}$ is open, the observer can observe the interference pattern. Conversely, when $\mathrm{BS_{out}}$ is closed, the observer can obtain the {\it which-way} information of each photon. The wave or particle nature of a photon is solely determined by the type of measurement that performed on it, despite the fact that the photon has already completed its travel. In the Wigner's S-G interferometer circuit, the determination or elimination of the {\it which-way} information of the qubit $q_{S}$ can be achieved by the application or removal of rotation gates $R_{y}(-\frac{\pi}{2})$, respectively. If a quantum random number generator (QRNG), which can be an ancilla qubit used to control the application or removal of gates $R_{y}(-\frac{\pi}{2})$, is activated only after $q_{S}$ has entered into the second S-G circuit, then we have successfully implemented the quantum circuit version of the delay-choice experiment. Compared to the optical realization \cite{delay1, delay2, delay3}, the implementation of the delayed choice operation in a circuit may offer advantages in terms of ease of realization. This is due to the ability to accurately determine the timing of gate operation in advance.
By applying $R_{y}(-\frac{\pi}{2})$, we obtain the which way information and the corresponding counting results are displayed in Fig. 3b. Conversely, removing $R_{y}(-\frac{\pi}{2})$ eliminates the {\it which-way} information, resulting in the counting results displayed in Fig. 3c. Similar to the optical experiment, the presence of odd state in Fig. 3c can be attributed to the introduction of a $\pi$ phase in the interferometer. 

In conclusion, we have presented the quantum circuits version of S-G devices, incorporating different measurement directions. We have provided a comprehensive analysis on the utilization of these devices for the implementation of sequential S-G experiments, specifically focusing on Wigner's S-G interferometer. The present study also explores the feasibility of executing Wheeler's delay-choice gedanken experiment by utilizing a sequential S-G quantum circuit.
The S-G quantum circuit is a parametric circuit with a shallow depth, consisting solely of nearest-neighbor two-qubits gates and single-qubit gates. This characteristic renders it highly suitable for practical implementation in the current NISQ hardwares. We propose that the optimal execution of sequential S-G quantum circuits can be achieved through the utilization of cross-shaped quantum processors. These processors can be readily implemented using today's superconducting qubit chips. We hold the belief that the utilization of quantum circuit version of the S-G device offers us a remarkable experimental platform for the purpose of showcasing and investigating the captivating facets of foundational quantum theory.  
More significantly, the launch of increasing advanced quantum computing cloud platforms that support a range of quantum hardware, such as BAQIS Quafu \cite{Quafu}, enables us to utilize cloud resources for the purpose of investigating the enigmatic aspects of quantum phenomena without the need for a specialized laboratory.

\hfill

\begin{acknowledgments}
\textbf{Acknowledgments:} Meng-Jun Hu expresses gratitude to Dong E. Liu, Peng Zhao and Heng Fan for their valuable discussions. Meng-Jun Hu is supported by Beijing Academy of Quantum Information Sciences. Hai-Xing Miao is supported by the State Key Laboratory of Low Dimensional Quantum Physics and the start-up fund provided by Tsinghua University. Yong-Sheng Zhang acknowledges the financial support provided is supported by the National Natural Science Foundation of China (Grant No. 92065113) and the Anhui Initiative in Quantum Information Technologies. This study aims to commemorate the centennial anniversary of the Stern-Gerlach experiment. 

{\bf Code available:} Relevant code pertaining to this study is accessible upon direct request to the corresponding author.
\end{acknowledgments}

\end{document}